\documentclass[aps, superscriptaddress, twocolumn, floatfix, letterpaper]{revtex4}
 
\usepackage{amssymb}
\usepackage{amsmath}
\usepackage{amsfonts}
\usepackage{upgreek}
\usepackage{mathrsfs}
\usepackage{bm}
\usepackage{bbold}
\usepackage{braket}
\usepackage{xspace}
\usepackage{graphicx}
\usepackage{comment}
\usepackage{hyperref}

\DeclareMathAlphabet\mathbfcal{OMS}{cmsy}{b}{n}
\newcommand{\mc}[1]{\mathcal{#1}}

\newcommand{\msf}[1]{\mathsf{#1}}

\allowdisplaybreaks

\newcommand{\eq}[1]{(\ref{#1})}
\newcommand{\Eq}[1]{Eq.~(\ref{#1})}
\newcommand{\Eqs}[1]{Eqs.~(\ref{#1})}
\newcommand{\Fig}[1]{Fig.~\ref{#1}}
\newcommand{\Sec}[1]{Sec.~\ref{#1}}

\newcommand{\Refs}[1]{Refs.~\onlinecite{#1}}
\newcommand{\App}[1]{Appendix~\ref{#1}}

\newcommand{\eg}{{e.g.,\/}\xspace}
\newcommand{\ie}{{i.e.,\/}\xspace}

\renewcommand{\Re}{\text{Re}\,}
\renewcommand{\Im}{\text{Im}\,}

\newcommand{\pd}{\partial}

\newcommand{\oper}[1]{\hat{#1}}
\newcommand{\favr}[1]{\langle #1 \rangle}
\newcommand{\cev}[1]{\reflectbox{\ensuremath{\vec{\reflectbox{\ensuremath{#1}}}}}}

\begin{document}

\title{Structure formation in turbulence as instability of effective quantum plasma}

\author{Vasileios Tsiolis}
\email{vtsiolis@princeton.edu}
\affiliation{Department of Astrophysical Sciences, Princeton University, Princeton, NJ 08544}

\author{Yao Zhou}
\affiliation{Princeton Plasma Physics Laboratory, Princeton, NJ 08543}

\author{I. Y. Dodin}
\affiliation{Department of Astrophysical Sciences, Princeton University, Princeton, NJ 08544}
\affiliation{Princeton Plasma Physics Laboratory, Princeton, NJ 08543}

\date{\today}

\begin{abstract}
Structure formation in turbulence is effectively an instability of ``plasma'' formed by fluctuations serving as particles. These ``particles'' are quantumlike; namely, their wavelengths are non-negligible compared to the sizes of background coherent structures. The corresponding ``kinetic equation" describes the Wigner matrix of the turbulent field, and the coherent structures serve as collective fields. This formalism is usually applied to manifestly quantumlike or scalar waves. Here, we extend it to compressible Navier--Stokes turbulence, where the fluctuation Hamiltonian is a five-dimensional matrix operator and diverse modulational modes are present. As an example, we calculate these modes for a sinusoidal shear flow and find two modulational instabilities. One of them is specific to supersonic flows, and the other one is a Kelvin--Helmholtz-type instability that is a generalization of the known zonostrophic instability. This work serves as a stepping stone toward improving the understanding of magnetohydrodynamic turbulence, which can be approached similarly.
\end{abstract}


\maketitle
\bibliographystyle{full.bst} 

\section{Introduction}
\label{sec:intro}

Turbulence is notoriously difficult to study theoretically, and \textit{ab~initio} simulations are often considered as the only feasible option. However, some aspects of turbulent dynamics, such as structure formation and modulational instabilities (MIs) in particular, can be made fairly intuitive by drawing analogies with plasma theory. To argue this, let us start with the following observation: any, even regular, dynamics of plasma can be viewed as turbulent dynamics of quantum matter waves. Kinetic theory of (non-degenerate) plasma hides the complexity of the quantum field by considering only the Fourier spectrum of its two-point correlation function, the Wigner function~$W$ \cite{ref:wigner32}, which satisfies the so-called Wigner--Moyal equation (WME) \cite{ref:moyal49}. In the classical limit, when the de~Broglie wavelengths and quantum correlations are negligible, $W$ can be interpreted as the particle distribution function. Accordingly, the WME becomes a Liouville-type equation, or the Vlasov equation in the collisionless limit \cite{book:landau2}, which is relatively intuitive and often manageable analytically.

A similar approach can be applied to classical wave turbulence, and indeed, Liouville-type `wave kinetic equations' (WKE) for inhomogeneous ensembles of nonlinear classical waves are widely known \cite{book:kadomtsev65}. However, since the wavelengths of classical fluctuations are typically much larger than those of quantum fluctuations, the geometrical-optics (GO) approximation underlying the WKE is more fragile than the classical limit of the kinetic theory of quantum plasma. For example, the WKE is often inadequate for modeling structure formation in classical wave turbulence. This is because the characteristic scales of the structures that form are often determined by diffraction, which the WKE neglects along with other full-wave effects and phase information in general \cite{foot:soliton}. Furthermore, unlike particle Hamiltonians, wave Hamiltonians are often not Hermitian even approximately (but may instead be \textit{pseudo}-Hermitian if there is no dissipation \cite{ref:larsson91, ref:brizard93, ref:brizard94, ref:qin19}; also see \Sec{sec:WM}). Because of this, practical applications of the WKE have been limited, and different theoretical formulations have been used instead, such as the so-called CE2 and alike \cite{phd:parker14, phd:squire15, ref:farrell03, ref:marston08, ref:tobias11, ref:srinivasan12, ref:bakas13, ref:tobias13, ref:constantinou14, ref:marston16}. However, these formulations are not intuitive, and their relation to the familiar GO limit is not obvious. Returning to the Wigner--Moyal formulation and applying it systematically to classical turbulence beyond the GO limit can fix these problems and thus is potentially advantageous.

The idea that classical turbulence can be described using quantumlike WMEs beyond the GO limit is recognized in literature to some extent. So far, it was successfully applied to \textit{manifestly quantumlike} systems, such as those governed by the nonlinear Schr\"odinger equation \cite{ref:hall02, ref:onorato03, ref:semenov08, ref:eliasson10, ref:hansson12, ref:hansson13, ref:picozzi14} and the Klein--Gordon equation \cite{ref:santos05, ref:santos07}. More recently, the same method was extended to drift-wave and Rossby-wave turbulence, where the wave function is governed by a Hamiltonian very different from a usual quantum particles, and a number of intriguing effects were identified as a result \cite{my:zonal, my:tertsum, my:ppo, my:soliton, my:wcol}. But this application is still limited to scalar waves, while the Wigner--Moyal approach could be useful also in more complex systems, where the wave function is a large-dimensional vector comprised of diverse fields (\ie not just the electromagnetic field, as usual). In this case, the derivation of the WME requires a more systematic approach which is yet to be worked out explicitly for typical turbulent systems.

Here, we explore an example of such system, namely, three-dimensional \textit{compressible} Navier-Stokes turbulence, which we assume inviscid for simplicity. This system is chosen because its governing equations are similar in form to those in many interesting physics problems, for example, magnetohydrodynamic (MHD) turbulence and turbulent dynamo \cite{phd:squire15, foot:tobias19}. Our goal is to develop a Wigner--Moyal formulation for Navier-Stokes turbulence as a stepping stone toward similar calculations for MHD, which are to be done in the future. Specifically, we derive a closed set of general equations which govern the turbulence in the `quasilinear approximation'. (This approximation essentially amounts to neglecting eddy--eddy interactions; see Secs.~\ref{sec:model} and~\ref{sec:example} for details.) As an example, we also apply these equations to derive the MIs of compressible shear flows, which happen to be tractable analytically. We find one stable modulational mode and two MIs. One of these MIs is specific to supersonic flows. The other one is a Kelvin--Helmholtz-type instability that is a generalization of the `zonostrophic' instability, which is well known for incompressible fluids \cite{ref:srinivasan12, phd:parker14, my:zonal, my:ppo}. This calculation is intended to demonstrate how the machinery of quantum statistical theory facilitates practical calculations in classical-turbulence theory by making them straightforward and systematic.

The paper is organized as follows. In \Sec{sec:model}, we introduce the basic governing equations and the quasilinear approximation. In \Sec{sec:WM}, we derive the Wigner--Moyal formulation. In \Sec{sec:modulational}, we outline the general calculation of the linear modulational dynamics for broadband homogeneous turbulence. In \Sec{sec:example}, we discuss a specific example. In \Sec{sec:conclusions}, we summarize the main results of our work. Auxiliary calculations are presented in appendixes. It is also to be noted that our calculations invoke the Weyl calculus, which is widely used in quantum theory and has been applied to classical waves too, albeit mainly in the linear regime. (For overviews, see, \eg Refs.~\onlinecite{ref:mcdonald88, book:tracy, phd:ruiz17}.) Readers who are not familiar with the Weyl calculus are encouraged to review the primer on this topic in \App{sec:weyl} before reading further.

\section{The model}
\label{sec:model}

We assume that the system is governed by the equations of inviscid hydrodynamics,
\begin{subequations}\label{eq:main}
\begin{gather}
\partial_t \mathbf{V} + \left(\mathbf{V}\!\cdot\! \nabla \right) \mathbf{V} = - \varrho^{-1}\nabla P,
\\
\partial_t \varrho + \nabla\!\cdot\!\left(\varrho \mathbf{V}\right) = 0,
\\
\left( \partial_t +  \mathbf{V}\!\cdot\!\nabla \right) (P \varrho^{-\gamma}) = 0,
\end{gather}
\end{subequations}
where $\mathbf{V}$ is the fluid velocity, $\varrho$ is the mass density, $P$ is the pressure, and $\gamma$ is some polytropic index which we assume to be constant. As a reference state, we assume a homogeneous background with zero velocity, constant density $\bar{\varrho}$, and constant pressure $\bar{P}$. Hence, we adopt
\begin{subequations}
\begin{gather}
\mathbf{V} = \mathbf{0} + \mathbf{v},\\
\varrho = (1 + n) \bar{\varrho},\\
P = (1 + \tau) \bar{P},
\end{gather}
\end{subequations}
where $\mathbf{v}$, $n$, and $\tau$ characterize deviations of the velocity, density, and pressure from the reference state. Assuming the notation $\favr{\ldots}$ for the statistical average, let us split these quantities as follows:
\begin{subequations}
\begin{gather}
\mathbf{v} = \bar{\mathbf{v}} + \tilde{\mathbf{v}},
\quad
\bar{\mathbf{v}} \doteq \favr{\mathbf{v}},\\
n = \bar{n} + \tilde{n}, 
\quad 
\bar{n} \doteq \favr{n},\\
\tau = \bar{\tau} + \tilde{\tau},
\quad
\bar{\tau} \doteq \favr{\tau},
\end{gather}
\end{subequations}
where the tilted quantities characterize the corresponding turbulent fluctuations and $\doteq$ denotes definitions. 

The equations for the barred variables are obtained by taking the average of \Eqs{eq:main}. We shall assume that terms of the fourth and higher order in the fluctuation amplitude can be neglected; hence, we obtain
\begin{align}
\partial_t & \bar{\mathbf{v}} + \left(\bar{\mathbf{v}}\!\cdot\!\nabla\right)\!\bar{\mathbf{v}} + \left\langle \left(\tilde{\mathbf{v}}\!\cdot\!\nabla\right)\!\tilde{\mathbf{v}} \right\rangle +  \notag\\
& \mbox{}\hspace{1cm}
+c^2(1+\bar{n})^{-1}\,(\!\nabla \bar{\tau}\!- \langle \tilde{N}\nabla \tilde{\tau}\rangle + \langle \tilde{N}^2\rangle\nabla\bar{\tau})\! = 0,
\notag
\\
\partial_t & \bar{n} + \bar{\mathbf{v}}\!\cdot\!\nabla \bar{n}  + \left(1 + \bar{n}\right)\nabla\!\cdot\!\bar{\mathbf{v}} + \left\langle  \tilde{\mathbf{v}}\!\cdot\!\nabla \tilde{n} \right\rangle + \left\langle \tilde{n}\nabla\!\cdot\!\tilde{\mathbf{v}}\right\rangle\!= 0, 
\notag
\\
\partial_t & \bar{\tau} + \bar{\mathbf{v}}\!\cdot\!\nabla\!\bar{\tau}  + \gamma \left(1\!+\!\bar{\tau}\right)\nabla\!\cdot\!\bar{\mathbf{v}} + \left\langle \tilde{\mathbf{v}}\!\cdot\!\nabla\tilde{\tau} \right\rangle +  \gamma \left\langle \tilde{\tau}\nabla\!\cdot\!\tilde{\mathbf{v} }\right\rangle\!=\!0,
\notag
\end{align}
where $c \doteq \sqrt{\bar{P}/\bar{\varrho}}$ is the isothermal sound speed and $\tilde{N}\doteq \tilde{n}/(1+\bar{n})$ is a rescaled fluctuation density.

By subtracting these from \Eqs{eq:main}, one obtains nonlinear equations for turbulent fluctuations. Although one can develop the Wigner--Moyal formulation for such nonlinear equations \cite{my:wcol}, below we adopt a simplified \textit{quasilinear} approach, in which the effect of nonlinearities in the equations for fluctuations is neglected. The quasilinear approach is known to adequately capture many effects in turbulence \cite{phd:parker14, phd:squire15, my:ppo, my:soliton, ref:srinivasan12}, and as to be shown in \Sec{sec:example}, this also extends to structure formation in Navier--Stokes fluids. Hence, the fluctuation equations are adopted in the following linearized form:
\begin{align}
& \partial_t \tilde{\mathbf{v}} + \left(\bar{\mathbf{v}}\!\cdot\!\nabla\right)\!\tilde{\mathbf{v}} + \left(\tilde{\mathbf{v}}\!\cdot\!\nabla\right)\!\bar{\mathbf{v}} +\notag\\
& \mbox{}\hspace{2cm}
+c^2(1+\bar{n})^{-1}\,(\nabla \tilde{\tau} - \tilde{N}\nabla\bar{\tau}) = 0,\notag
\\
& \partial_t \tilde{n} + \bar{\mathbf{v}}\!\cdot\!\nabla \tilde{n} + \left(1 + \bar{n}\right)\nabla\!\cdot\!\tilde{\mathbf{v}} + \tilde{\mathbf{v}}\!\cdot\!\nabla \bar{n} + \tilde{n}\nabla\!\cdot\!\bar{\mathbf{v}} = 0,\notag
\\
& \partial_t \tilde{\tau} + \bar{\mathbf{v}}\!\cdot\!\nabla\tilde{\tau} + \tilde{\mathbf{v}}\!\cdot\!\nabla \bar{\tau} + \gamma \tilde{\tau} \nabla\!\cdot\!\bar{\mathbf{v}} + \gamma \left(1+\bar{\tau}\right)\nabla\!\cdot\!\tilde{\mathbf{v}} = 0.
\notag
\end{align}
One can rewrite them more compactly in terms of
\begin{gather}
\tilde{\psi} \doteq \begin{pmatrix}
\tilde{\mathbf{v}} \\
\tilde{n} \\
\tilde{\tau}
\end{pmatrix},
\label{eq:intro_column_vector}
\end{gather}
which is a five-dimensional vector, since $\tilde{\mathbf{v}}$ is three-dimensional. (Note that the first three rows of $\tilde{\psi}$ have the dimension of velocity, whereas the fourth and fifth rows are dimensionless.) Specifically, one can write the fluctuation equations as a vector Schr\"odinger equation
\begin{gather}\label{eq:schr}
i\partial_t \tilde{\psi} = \hat{H}\tilde{\psi},
\end{gather}
so $\tilde{\psi}$ can be viewed as the state function of an effective quantum particle. (Since $\tilde{\psi}$ is a vector, this particle can be assigned a spin, as described in Ref.~\onlinecite{my:qdirac}.) Here, $\hat{H} = \hat{H}_0 + \hat{H}_{\rm int}$ is a linear operator serving as a (non-Hermitian) Hamiltonian. The part $\hat{H}_0$ is entirely determined by the reference state; namely, $\hat{H}_0 \!=\! H_0(\oper{\mathbf{k}})$, where
\begin{gather}
H_0(\mathbf{k}) = \begin{pmatrix}
 0   &   0   &  c^2 {\mathbf{k}} \\
 {\mathbf{k}}\cdot & 0 & 0 \\
 \gamma  {\mathbf{k}}\cdot & 0 & 0
 \end{pmatrix},
\label{eq:H0}
\end{gather}
$\hat{\mathbf{k}} \doteq - i\nabla$ is the wave-vector operator, and $\cdot$ denotes the scalar product, as usual. The part $\hat{H}_{\rm int}$ is determined by the regular perturbations $\lbrace\bar{\mathbf{v}}, \bar{n}, \bar{\tau}\rbrace$ to the reference state and is given by
\begin{gather}
\mbox{}\kern-5pt
    \hat{H}_{\rm int} \!=\! \begin{pmatrix}
    \mathbb{1}_3(\bar{\mathbf{v}}\!\cdot\!\hat{\mathbf{k}}) + \boldsymbol{\mc{X}} & ic^2(1\!+\!\bar{n})^{-2}\nabla \bar{\tau} & -c^2\bar{n}(1\!+\!\bar{n})^{-1}\hat{\mathbf{k}} \\
   (\bar{n}\hat{\mathbf{k}} \!-\!i\nabla \bar{n}) \cdot & \bar{\mathbf{v}}\!\cdot\!\hat{\mathbf{k}}\!-\!i\nabla\!\cdot\! \bar{\mathbf{v}}  & 0 \\
    (\gamma \bar{\tau}\hat{\mathbf{k}}\!-\!i\nabla \bar{\tau}) \cdot & 0 & \bar{\mathbf{v}}\!\cdot\!\hat{\mathbf{k}} \!-\! i\gamma\nabla\!\cdot\!\bar{\mathbf{v}}
    \end{pmatrix}\!\!,\notag
\end{gather}
where $\mathbb{1}_N$ is a $N \times N$ unit matrix, $\mc{X}_{ab} \doteq -i\partial_b \bar{v}_a$, and $\pd_b \doteq \partial/\partial x_b$. For a more explicit representation of $\hat{H}_0$ and $\hat{H}_{\rm int}$, see \Eqs{eq:AppB_H0} and~\eq{eq:AppB_Hint} in \App{app:matrices}.

\section{The Wigner--Moyal equation}
\label{sec:WM}

\subsection{Basic equations}

Let us consider a family of all reversible linear transformations of $\tilde{\psi}(t, \mathbf{x})$, which map $\tilde{\psi}(t, \mathbf{x})$ into some family of image functions. Since these functions are mutually equivalent up to an isomorphism, the resulting family can be viewed as a single object, a time-dependent `state vector' $\ket{\tilde{\psi}}$. The original function $\tilde{\psi}(t, \mathbf{x})$ can then be understood as a projection of $\ket{\tilde{\psi}}$, namely, as its `coordinate representation' given by $\tilde{\psi}(t, \mathbf{x}) = \braket{\mathbf{x}|\tilde{\psi}}$. Here, $\ket{\mathbf{x}}$ are the eigenstates of the position operator $\hat{\mathbf{x}}$ normalized such that $\braket{\mathbf{x}'| \hat{\mathbf{x}}|\mathbf{x}} = \mathbf{x}\braket{\mathbf{x}'|\mathbf{x}} = \mathbf{x}\mathrm{\delta} (\mathbf{x}'-\mathbf{x})$. This definition of a field is similar to that used in quantum mechanics for describing probability amplitudes; hence, it is convenient to describe the dynamics of $\ket{\tilde{\psi}}$ using a quantumlike formalism. This is implemented as follows.

Consider the `density operator' $\hat{W} \doteq \ket{\tilde{\psi}}\bra{\tilde{\psi}}$. From the abstract vector form of \Eq{eq:schr}, which is
\begin{gather}\label{eq:schrabs}
i\partial_t \ket{\tilde{\psi}} = \hat{H}\ket{\tilde{\psi}},
\end{gather}
one then obtains the following operator equation:
\begin{gather}\label{eq:vN}
i\partial_t \hat{W} = \hat{H}\hat{W} - \hat{W}\hat{H}^\dag,
\end{gather}
which can be understood as a generalized von Neumann equation. The term `generalized' refers to the fact that unlike a typical quantummechanical Hamiltonian, $\hat{H}$ is not Hermitian; instead, it is \textit{pseudo}-Hermitian \cite{ref:mostafazadeh02}, as is our original system.  As the next step, we project \Eq{eq:vN} on the phase space $(\mathbf{x}, \mathbf{k})$ using the Wigner--Weyl transform (\App{sec:weyl}), just like it is done in the statistical approach to quantum mechanics \cite{ref:moyal49, ref:groenewold46, ref:mendonca11b, ref:mendonca12}. Then, \Eq{eq:vN} leads to the WME \cite{ref:mcdonald88},
\begin{gather}
 i\partial_t W = H \star W - W \star H^\dag,
 \label{eq:WM}
\end{gather}
where $\star$ is the `Moyal product' \eq{eq:AppendixA_Moyal_Star_Product}, the matrix $H$ is the Weyl symbol of $\hat{H}$, and the matrix $W$ is the Weyl symbol of $\hat{W}$. The matrix $W$ is also known as the Wigner function \cite{ref:wigner32}, or more precisely, the \textit{Wigner matrix} in our case, and its elements can be expressed as follows:
\begin{gather}
W_{\alpha \beta}(t, \mathbf{x}, \mathbf{k}) \doteq\! \mathrm{\int} \mathrm{d}^3s\  e^{-i\mathbf{k}\cdot\mathbf{s}}\tilde{\psi}_\alpha \left(t, \mathbf{x}_{+}\right)\tilde{\psi}_\beta^*\left(t, \mathbf{x}_{-} \right),
\notag
\end{gather}
where $\mathbf{x}_{\pm} = \mathbf{x} \pm \mathbf{s}/2$; hence, $W$ is Hermitian, i.e.,
\begin{gather}\label{eq:Wherm}
W_{\alpha \beta}(t, \mathbf{x}, \mathbf{k}) = W^*_{\beta \alpha}(t, \mathbf{x}, \mathbf{k}).
\end{gather}
(The Greek indices $\alpha$ and $\beta$ are assumed to span from~1 to~5, as opposed to the Latin indices $a$ and $b$ used above and below, which are assumed to span from~1 to~3.) Also note that in our case, the field vector $\tilde{\psi}(t, \mathbf{x}) \doteq \braket{\mathbf{x}|\tilde{\psi}}$ is real, so the Wigner matrix satisfies
\begin{gather}\label{eq:Wsym}
W_{\alpha \beta}(t, \mathbf{x}, \mathbf{k}) = W_{\beta \alpha}(t, \mathbf{x}, -\mathbf{k}).
\end{gather}

Let us also introduce the statistical average of the Wigner matrix, $\bar{W} \doteq \favr{W}$. Its elements are given by
\begin{gather}
\bar{W}_{\alpha\beta}(t, \mathbf{x}, \mathbf{k}) \!\doteq \!\!\mathrm{\int} \mathrm{d}^3s\  e^{-i\mathbf{k}\cdot\mathbf{s}} \favr{\tilde{\psi}_\alpha \left(t, \mathbf{x}_{+}\right)\tilde{\psi}_\beta^*\left(t, \mathbf{x}_{-} \right)},
\end{gather}
so $\bar{W}$ can be interpreted as the spatial spectrum of the autocorrelation matrix corresponding to $\tilde{\psi}$. Also note that since $H$ is independent of fluctuations, the equation for $\bar{W}$ is readily obtained by averaging the WME \eq{eq:WM},
\begin{gather}
i\partial_t \bar{W} = H \star \bar{W} - \bar{W} \star H^\dag.
\label{eq:WM3}
\end{gather}
(Beyond the quasilinear approximation, beatings of the fluctuating parts of $W$ and $H$ give rise to an additional term which serves as a `collision operator' \cite{my:wcol}.)

Note that using \Eq{eq:AppendixA_Moyal_Star_Product}, one can write
\begin{subequations}\label{eq:moyalstar}
\begin{gather}
H \star \bar{W} = H \left[\sum_{s = 0}^\infty \left(\frac{i}{2}\,\hat{\mc{L}}\right)^s\right] \bar{W},\\
\bar{W} \star H^\dag = \bar{W} \left[\sum_{s = 0}^\infty \left(\frac{i}{2}\,\hat{\mc{L}}\right)^s\right] H^\dag,
\end{gather}
\end{subequations}
where $\hat{\mc{L}} \doteq \cev{\partial_{\mathbf{x}}}\cdot\vec{\partial_{\mathbf{k}}} - \cev{\partial_{\mathbf{k}}}\cdot\vec{\partial_{\mathbf{x}}}$ and the arrows indicate the directions in which the derivatives act. If both $\bar{W}$ and $H$ are sufficiently smooth functions of $(\mathbf{x}, \mathbf{k})$ (and $t$), with scales $\Delta x$ and $\Delta k$ satisfying $\epsilon^{-1} \doteq (\Delta x)(\Delta k) \gg 1$, then $\hat{\mc{L}} \sim \epsilon \ll 1$. In this case, which corresponds to the GO limit, the above series can be replaced with just the first two terms. This leads to an equation similar to a quasilinear WKE or the Vlasov equation for classical plasma,
\begin{gather}\label{eq:wke}
\partial_t \bar{W} =
\lbrace H_H, \bar{W} \rbrace + 2(H \bar{W})_A.
\end{gather}
Here, $\lbrace \ldots , \ldots  \rbrace$ denotes the canonical Poisson bracket [\Eq{eq:AppendixA_Poisson_bracket}], and the indices $_H$ and $_A$ denote, respectively, the Hermitian and anti-Hermitian parts of a given matrix; namely, $M_H \doteq (M + M^\dagger)/2$ and $M_A \doteq (M - M^\dagger)/(2i)$ for any~$M$. We shall \textit{not} rely on the approximation \eq{eq:wke}, because it misses essential physics; for example, as to be argued in \Sec{sec:example}, the MI's maximum growth rate cannot be predicted from \Eq{eq:wke}. Hence, retaining the high-order terms in \Eqs{eq:moyalstar} is in fact essential, and the general `quantumlike' kinetic equation \eq{eq:WM3} is preferred over its GO limit \eq{eq:wke}.

\subsection{The explicit form of the dynamic equations}

The symbol $H$ that enters \Eq{eq:WM3} can be expressed as $H = H_0 + H_{\rm int}$, where $H_0$ is given by \Eq{eq:H0} and $H_{\rm int}$ is presented in \App{app:matrices}. In order to close the system, let us express the equations for the statistically-averaged fields $\lbrace\bar{\mathbf{v}}, \bar{n}, \bar{\tau}\rbrace$ through $\bar{W}$. The turbulent terms that appear in the equations for these fields can be rewritten in terms of $\bar{W}$ as follows. For example,
\begin{align}
\left[ \left(\tilde{\mathbf{v}}\cdot\nabla\right)\tilde{\mathbf{v}}\right]_a &= i\tilde{v}_b\hat{k}_b \tilde{v}_a  \nonumber \\
& =  i\braket{\mathbf{x}|\tilde{v}_b} \braket{\mathbf{x}|\hat{k}_b | \tilde{v}_a} \nonumber\\
& =  i \braket{\mathbf{x}|\hat{k}_b | \tilde{v}_a}\braket{\tilde{v}_b|\mathbf{x}} \nonumber\\
& =  i \braket{\mathbf{x}|\hat{k}_b \oper{W}_{ab} |\mathbf{x}},
\end{align}
where summation over repeating Latin indices is assumed from~1 to~3. (We shall also use index $n$ for~4 and index $\tau$ for~5.) Using also \Eq{eq:AppendixA_Weyl_symbol_of_C}, we then obtain an explicit form of this and other similar quantities: 
\begin{subequations}\label{eq:variouststars}
\begin{gather}
\left[ \left(\tilde{\mathbf{v}}\cdot\nabla\right)\tilde{\mathbf{v}}\right]_a 
= i \int \dfrac{\mathrm{d}^3k}{(2\pi)^3}\, k_b \star W_{a b},
\\
\left(\tilde{n}\nabla \tilde{\tau}\right)_a 
= i \int \dfrac{\mathrm{d}^3k}{(2\pi)^3}\,k_a \star W_{\tau n},
\\
\tilde{n}^2 
= \int \dfrac{\mathrm{d}^3k}{(2\pi)^3}\, W_{n n},
\\
\tilde{\mathbf{v}}\cdot\nabla \tilde{n} 
= i\int \dfrac{\mathrm{d}^3k}{(2\pi)^3}\, k_b \star W_{n b},
\\
\tilde{n}\nabla\cdot\tilde{\mathbf{v}} = 
i \int\dfrac{\mathrm{d}^3k}{(2\pi)^3} k_b \star W_{bn},
\\
\tilde{\mathbf{v}}\cdot\nabla \tilde{\tau} 
= i \int \dfrac{\mathrm{d}^3 k}{(2\pi)^3}\, k_b \star W_{\tau b},
\\
\tilde{\tau} \nabla\cdot \tilde{\mathbf{v}} = i \int \dfrac{\mathrm{d}^3k}{(2\pi)^3}\,k_b \star W_{b\tau}.
\end{gather}
\end{subequations}
Using \Eqs{eq:variouststars} and the equations derived in \Sec{sec:model}, we obtain: 
\begin{widetext}
\begin{subequations}\label{eq:averagedPsiEqs}
\begin{gather}
\partial_t \bar{v}_a + \bar{v}_b \pd_b \bar{v}_a + \dfrac{c^2}{1+\bar{n}} \pd_a \bar{\tau} + i\int\dfrac{\mathrm{d}^3 k}{(2\pi)^3} \left[-
\frac{c^2}{(1 + \bar{n})^2}\, k_a \star \bar{W}_{\tau n}
-\frac{ic^2 \pd_a \bar{\tau} }{(1 + \bar{n})^3}\, \bar{W}_{nn}
+ k_b \star \bar{W}_{ab}
\right] = 0, \label{eq:Averaged_velocity_W}\\
\partial_t \bar{n} + \bar{v}_b \pd_b \bar{n} + \left(1+\bar{n}\right) \pd_b \bar{v}_b + i\int\dfrac{\mathrm{d}^3k}{(2\pi)^3} \,
k_b \star (\bar{W}_{nb} + \bar{W}_{bn}) = 0, 
\label{eq:Averaged_density_W}\\
 \partial_t \bar{\tau} + \bar{v}_b \pd_b \bar{\tau} + \gamma\left(1+\bar{\tau}\right) \pd_b \bar{v}_b + i\int \dfrac{\mathrm{d}^3k}{(2\pi)^3}\, 
k_b \star (\gamma\bar{W}_{b\tau} + \bar{W}_{\tau b}) = 0.
 \label{eq:Averaged_pressure_W}
\end{gather}
\end{subequations}
\end{widetext}
Also notice that the integrals here are manifestly imaginary due to \Eqs{eq:Wherm} and \eq{eq:Wsym}; hence, $i k_a \star W_{\alpha\beta}$ can as well be replaced with $-\Im (k_a \star W_{\alpha\beta})$, which is real.

Equations \eq{eq:averagedPsiEqs} contain the same physics as those one one would obtain within the CE2 approximation; however, they are more intuitive in that they are similar in form to the equations of effective collisionless quantum plasma. Specifically, if turbulent fluctuations are considered as effective vector particles with phase-space coordinates $(\mathbf{x}, \mathbf{k})$, then $\bar{W}$ can be viewed as the `particle' distribution in phase space, $H$ can be viewed as the `particle' Hamiltonian, and 
\begin{gather}
\bar{\psi} \doteq \begin{pmatrix}
\bar{\mathbf{v}} \\
\bar{n} \\
\bar{\tau}
\end{pmatrix}
\label{eq:barpsi}
\end{gather}
serves as a collective vector field through which the `particles' interact. Although $\bar{W}$ and $H$ are matrices rather than scalars, collective oscillations in such `plasma' can be studied in the same way as collective oscillations are studied in usual plasmas. As a special case, below we consider linear waves, which correspond to weak modulational oscillations (instabilities) of the original system.

\section{Dynamics of weak modulations}
\label{sec:modulational}

\subsection{Basic equations}

Let us consider linear modulational dynamics of this system. We assume
\begin{gather}
\bar{W} = (2\pi)^3 \big[ 
F(\mathbf{k}) + f(t, \mathbf{x}, \mathbf{k})
\big],
\end{gather}
where $F(\mathbf{k})$ describes some equilibrium homogeneous turbulent state as a background and $f \ll F$ is a small inhomogeneous perturbation. We also assume that $\bar{\psi}$ is of order~$f$. Then, by linearizing \Eq{eq:WM3}, the following equation for the perturbation~$f$ is obtained:
\begin{gather}
i\partial_t f = 
H_0\star f - f \star H_0^\dag 
+ h \star F - F\star h^\dag,
\label{eq:WMlin}
\end{gather}
where $h$ is the linearized $H_{\rm int}$ (\App{app:matrices}). Also, the linearized \Eqs{eq:averagedPsiEqs} are
\begin{subequations}\label{eq:averagedPsiEqsLinearized}
\begin{align}
& \partial_t \bar{v}_a + c^2 \pd_a \bar{\tau} + i\int \mathrm{d}^3 k \big[
2c^2\bar{n} (k_a \star F_{\tau n})
\notag \\
& \mbox{}\hspace{.6cm}- ic^2 (\pd_a \bar{\tau})  F_{nn}
- c^2 k_a \star f_{\tau n} 
+ k_b \star f_{ab}
\big] = 0,\\
& \partial_t \bar{n} + \pd_b \bar{v}_b + i\int \mathrm{d}^3 k \,
k_b \star (f_{nb} + f_{bn}) = 0, 
\\
& \partial_t \bar{\tau} + \gamma \pd_b \bar{v}_b + i\int \mathrm{d}^3 k\, 
k_b \star (\gamma f_{b\tau} + f_{\tau b}) = 0.
\end{align}
\end{subequations}
The star products can be simplified using \Eq{eq:AppA_k_star_A}, and \Eqs{eq:averagedPsiEqsLinearized} eventually become
\begin{subequations}\label{eq:averagedPsiEqsLinearized2}
\begin{align}
& \partial_t \bar{v}_a + c^2 \pd_a \bar{\tau} + i\!\int\mathrm{d}^3 k\, \big[
2c^2\bar{n} k_a F_{\tau n}
-i\ c^2(\partial_a \bar{\tau})F_{nn} \notag \\
& \mbox{}\hspace{10pt} - c^2 (k_a f_{\tau n} - \dfrac{i}{2} \partial_a f_{\tau n}) + k_b f_{ab} - \dfrac{i}{2} \partial_b f_{ab} \big] = 0,  \\
& \partial_t \bar{n} + \pd_b \bar{v}_b + i\int \mathrm{d}^3 k \,
\big( k_b f_{nb} - \dfrac{i}{2} \partial_b f_{nb} \notag \\
& \mbox{}\hspace{2.6cm} + k_b f_{bn} - \dfrac{i}{2} \partial_b f_{bn} \big)   = 0, 
\\
& \partial_t \bar{\tau} + \gamma \pd_b \bar{v}_b + i\int \mathrm{d}^3 k\, 
\big[\gamma (k_b f_{b\tau} - \dfrac{i}{2} \partial_b f_{b\tau}) \notag \\
& \mbox{}\hspace{2.6cm} + k_b f_{\tau b} - \dfrac{i}{2} \partial_b f_{\tau b} \big] = 0.
\end{align}
\end{subequations}
Also note that this set of equations can be considered as a vector equation for the collective field $\bar{\psi}$ [\Eq{eq:barpsi}],
\begin{gather}\label{eq:barpsieq}
(i\partial_t - \oper{\mc{H}})\bar{\psi} = \int\mathrm{d}^3 k\,\oper{\Pi}f,
\end{gather}
where the operators $\oper{\mc{H}}$ and $\oper{\Pi}$ are easily determined from \Eqs{eq:averagedPsiEqsLinearized2}.

\subsection{Linear eigenmodes}

Let us assume the following complex notation:  
\begin{subequations}
\begin{gather}
f = (\msf{f}e^{i\Theta})_H,
\quad
\bar{\mathbf{v}} = \Re (\boldsymbol{\msf{v}}e^{i\Theta}),
\\
\bar{n} = \Re (\msf{n} e^{i\Theta}),
\quad
\bar{\tau} = \Re (\uptau e^{i\Theta}),
\end{gather}
\end{subequations}
or more compactly,
\begin{gather}
\bar{\psi} = \Re\,(\xi e^{i\Theta}),
\end{gather}
where $\Theta$ is the modulation phase. (Note that $\msf{f}$ is a $5 \times 5$ matrix, $\boldsymbol{\msf{v}}$ is a vector, and $\msf{n}$ and $\uptau$ are scalars.) Explicitly,
\begin{gather}
f = \dfrac{1}{2}\,(\msf{f}e^{i\Theta} +\ \msf{f}^{\dagger}e^{-i\Theta}),
\quad
\bar{\mathbf{v}} = \dfrac{1}{2}\,(\boldsymbol{\msf{v}}e^{i\Theta} + \,\boldsymbol{\msf{v}}^{*}e^{-i\Theta}),
\notag
\\
\bar{n} = \dfrac{1}{2}\,(\msf{n}e^{i\Theta} + \msf{n}^* e^{-i\Theta}),
\quad
\bar{\tau} = \dfrac{1}{2}\,(\uptau e^{i\Theta} + \uptau^* e^{-i\Theta}).
\notag
\end{gather}
We also adopt
\begin{gather}
h = \dfrac{1}{2}\,\big[\msf{h}^{(+)} e^{i\Theta} +\msf{h}^{(-)\dag}e^{-i\Theta}\big],
\label{eq:H_h1_with_F_and_G}
\end{gather}
where $\msf{h}^{(+)}$ and $\msf{h}^{(-)}$ are $5 \times 5$ matrices given by \Eqs{eq:AppB_h+} and~\eq{eq:AppB_h-}. Assuming $\Theta = - \Omega t + \mathbf{K} \cdot \mathbf{x}$, where the frequency $\Omega$ and the wave vector $\mathbf{K}$ are constant, \Eqs{eq:averagedPsiEqsLinearized2} take the following form:
\begin{subequations}\label{eq:OmKgeneral}
\begin{align}
& \Omega \msf{v}_a - c^2 K_a \uptau - \int\mathrm{d}^3 k\, 
(2c^2\msf{n} k_a F_{\tau n} + c^2 \uptau K_a F_{nn})\notag\\
& \mbox{}\hspace{2pt} =\! \int\mathrm{d}^3 k\, \big[\!-\!(k_a \!+\! K_a/2) c^2\msf{f}_{\tau n} + (k_b + K_b/2) \msf{f}_{ab} \big],  \\
& \Omega \msf{n} - K_b \msf{v}_b = \int\mathrm{d}^3 k \,
(k_b + K_b/2)(\msf{f}_{nb} + \msf{f}_{bn}), 
\\
& \Omega \uptau - \gamma K_b \msf{v}_b = \int \mathrm{d}^3 k\, 
(k_b + K_b/2) (\gamma\msf{f}_{b\tau} + \msf{f}_{\tau b}).
\end{align}
\end{subequations}
It follows that \Eq{eq:barpsieq} becomes
\begin{gather}\label{eq:barpsieqwk}
[\Omega - \mc{H}(\mathbf{K})]\xi = \int\mathrm{d}^3 k\, \Pi(\mathbf{K})\msf{f},
\end{gather}
where $\mc{H}$ is the symbol of $\oper{\mc{H}}$ and $\Pi$ is the symbol of $\oper{\Pi}$. (The dependence on $\mathbf{k}$ is not emphasized but assumed.) Also, \Eq{eq:WMlin} becomes the following matrix equation:
\begin{gather}
A\msf{f} + \msf{f}B = C.
\label{eq:matrix_equation_for_W}
\end{gather}
Here, we introduced
\begin{subequations}
\begin{gather}
A = \mathbb{1}_5\Omega/2 - \bar{H}_{0}\left(\mathbf{k}+\mathbf{K}/2\right),\\
B = \mathbb{1}_5\Omega/2 + \bar{H}_{0}^{\dagger}\left(\mathbf{k}-\mathbf{K}/2\right),\\
C = \msf{h}^{(+)} F \left(\mathbf{k}-\mathbf{K}/2\right) - F\left(\mathbf{k}+\mathbf{K}/2\right)\msf{h}^{(-)},
\end{gather}
\end{subequations}
and $\mathbb{1}_5$ is a $5\times 5$ unit matrix.

Since $\msf{h}^{(+)}$ and $\msf{h}^{(-)}$ depend on $\xi$, one can use \Eq{eq:matrix_equation_for_W} to express $\msf{f}$ as a linear function of $\xi$. Then, $\int\mathrm{d}^3 k\,\Pi \msf{f} = \Xi\xi$, where $\Xi$ is some $5 \times 5$ matrix. One can substitute this result into \Eq{eq:barpsieqwk} to obtain a vector equation for the linear modulational modes, ${\mc{Q}\left(\Omega, \mathbf{K}\right)\xi = 0}$, where
\begin{gather}\label{eq:barpsieqwk2}
\mc{Q}\left(\Omega, \mathbf{K}\right)
\doteq 
\Omega \mathbb{1}_5 \!-\! \mc{H}(\mathbf{K}) \!-\! \Xi(\Omega, \mathbf{K}).
\end{gather}
Accordingly, the dispersion relation of these modes is
\begin{gather}\label{eq:Qdet0}
\det \mc{Q}\left(\Omega, \mathbf{K}\right) = 0.
\end{gather}

As a side remark, note that the quasilinear dispersion relation \eq{eq:Qdet0} for modulational modes can be obtained automatically using computer algebra \cite{foot:mathematica} for any set of governing equations $D(\mathbf{\psi}, \pd\mathbf{\psi}) = 0$ [here, \Eqs{eq:main}], where $D$ is some nonlinear matrix function, $\psi$ is some real vector field on spacetime [here, $\mathbf{\psi} = \lbrace V_x, V_y, V_z, \varrho, P\rbrace$], and $\pd$ is the spacetime derivative. In case of complex $\psi$, the same formalism applies too, except one may need to treat $\Re\psi$ and $\Im\psi$ as separate fields.

\begin{figure}
\begin{center}
\includegraphics[width=.48\textwidth]{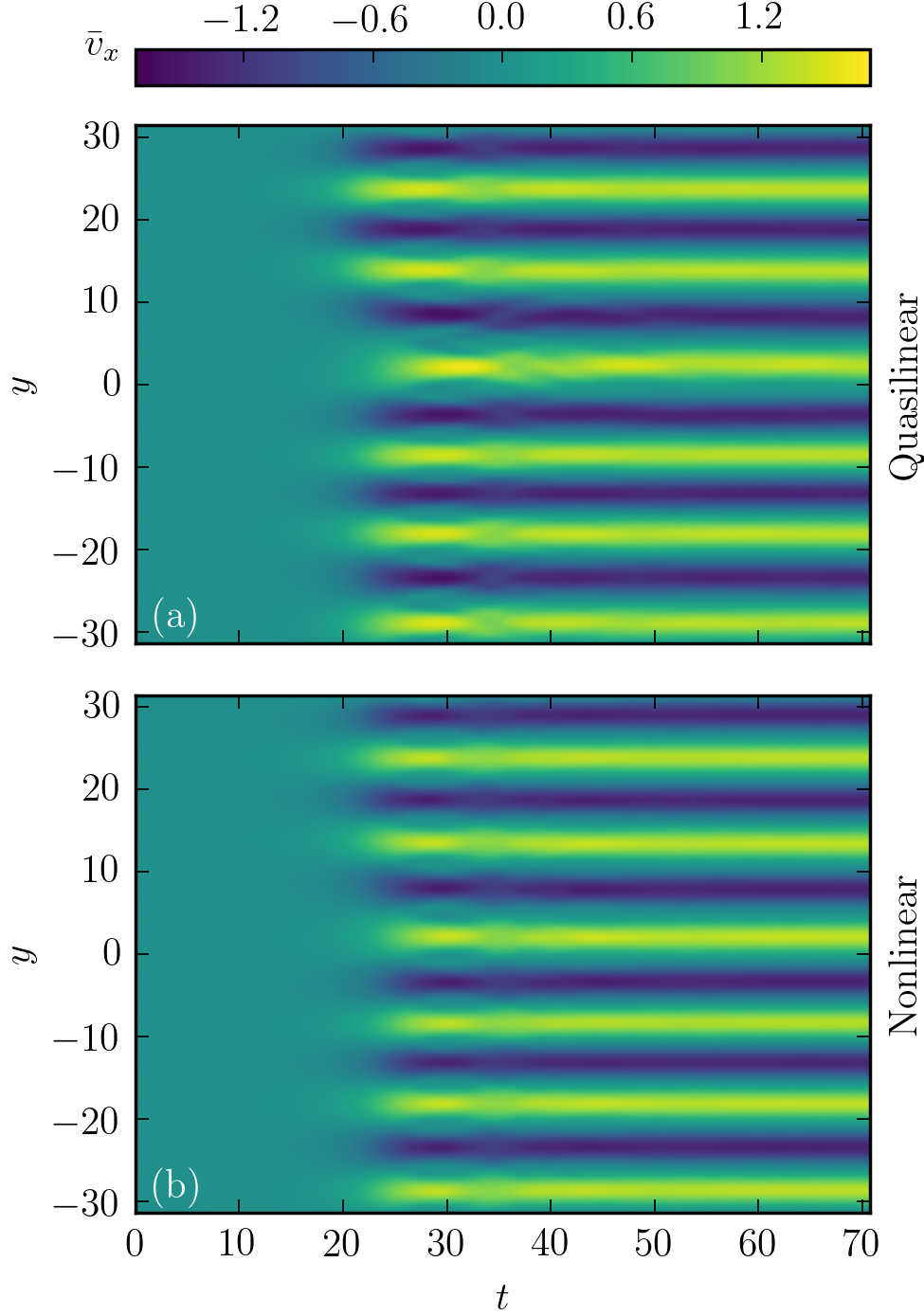}
\caption{Two-dimensional numerical simulations of a MI of a sinusoidal incompressible (${\gamma \to \infty}$) shear flow, with small hyper-viscosity added for numerical stability: (a)~quasilinear simulation, (b)~nonlinear simulation. Both simulations are initialized with identical random perturbations on top of the initial equilibrium \eq{eq:iniprofile}. The colormap shows the spatial-temporal evolution of $\bar{v}_x$; $\bar{v}_x$ is measured in units $\mc{U}$, $y$ is measured in units $k_x^{-1}$, and $t$ is measured in units $(k_x \mc{U})^{-1}$.}
\label{fig:dns}
\end{center}
\end{figure} 

\section{Example}
\label{sec:example}

\subsection{Basic equations}

As an example, we consider the MI of a shear flow with homogeneous density and temperature and the initial-velocity profile
\begin{gather}\label{eq:iniprofile}
\tilde{\mathbf{v}} = \mathbf{e}_y \,\mc{U}\sqrt{2}\cos(k_x x).
\end{gather}
(We assume $\mc{U} > 0$ and $k_x > 0$ for clarity. Also, $\mathbf{e}_y$ is the unit vector along the $y$ axis.) The corresponding matrix $F$ has only one nonzero element, 
\begin{gather}\label{eq:Fyy}
F_{yy}(\mathbf{k}) = \mc{U}^2 \delta (k_z)\delta(k_y)\left[ \delta\left(k-k_x \right) + \delta \left(k + k_x\right) \right]/2.
\end{gather}
Also, let us assume for simplicity that
\begin{gather}\label{eq:Ky}
\mathbf{K} = \mathbf{e}_y K_y.
\end{gather}
For this system, the quasilinear approximation invoked in Sec.~\ref{sec:model}-\ref{sec:modulational} adequately captures the linear modulational dynamics, as seen from direct numerical simulations (\Fig{fig:dns}). Hence, we can use \Eqs{eq:OmKgeneral}, which can now be simplified as follows:
\begin{subequations}
\begin{align}
& \Omega \msf{v}_x 
= \int \mathrm{d}^3k \big[-c^2 k_x \msf{f}_{\tau n} + k_x \msf{f}_{xx}  
\notag \\ 
& \mbox{}\hspace{15pt} + (k_y + K_y/2)\msf{f}_{xy} + k_z\msf{f}_{xz}], 
\\ 
& \Omega \msf{v}_y - c^2 K_y \uptau 
=\! \int\mathrm{d}^3 k\, \big[\!-\!(k_y \!+\! K_y/2) c^2\msf{f}_{\tau n}  
\notag \\
& \mbox{}\hspace{15pt} +\ k_x \msf{f}_{yx} + (k_y + K_y/2) \msf{f}_{yy} + k_z \msf{f}_{yz} \big], 
\\
& \Omega \msf{v}_z 
= \int \mathrm{d}^3k \big[-c^2 k_z \msf{f}_{\tau n} + k_x \msf{f}_{zx}  
\notag \\ 
& \mbox{}\hspace{15pt} + (k_y + K_y/2)\msf{f}_{zy} + k_z\msf{f}_{zz}\big], 
\\ 
& \Omega \msf{n} - K_y \msf{v}_y 
= \int\mathrm{d}^3 k \,\big[k_x(\msf{f}_{nx} + \msf{f}_{xn}) 
\notag \\
& \mbox{}\hspace{15pt} +(k_y + K_y/2)(\msf{f}_{ny} + \msf{f}_{yn}) + k_z(\msf{f}_{nz} + \msf{f}_{zn})\big], 
\\
& \Omega \uptau - \gamma K_y \msf{v}_y 
= \int \mathrm{d}^3 k\ \big[ k_x (\gamma \msf{f}_{x\tau} + \msf{f}_{\tau x})  
\notag \\
& \mbox{}\hspace{15pt} + (k_y + K_y/2)(\gamma \msf{f}_{y\tau} + \msf{f}_{\tau y}) + k_z (\gamma \msf{f}_{z\tau} + \msf{f}_{\tau z})\big]. 
\end{align}
\end{subequations}
We then calculate $\msf{f}$ using \Eq{eq:matrix_equation_for_W}. Next, using \Eq{eq:barpsieqwk2}, we also calculate $\mc{Q}(\Omega, \mathbf{K})$, which is found to be the following matrix \cite{foot:mathematica}:
\begin{widetext}
\begin{gather}
\left(
\begin{array}{ccccc}
\Omega + \frac{K_y^2 \mc{U}^2 [\Omega^2 -(K_y^2 - k_x^2) \mc{C}^2]}{\Omega[(k_x^2+K_y^2) \mc{C}^2 -\Omega^2]} 
& \frac{k_x K_y^3 \mc{C}^2 \mc{U}^2}{\Omega[(k_x^2+K_y^2) \mc{C}^2 -\Omega^2]} 
& 0 
& 0 
& \frac{k_x K_y^2 \mc{C}^2 \mc{U}^2}{\gamma[(k_x^2+K_y^2) \mc{C}^2-\Omega^2]} 
\medskip\\
\frac{2 k_x K_y \mc{U}^2 \Omega}{(k_x^2+K_y^2) \mc{C}^2-\Omega^2} 
& \Omega +\frac{K_y^2 \mc{U}^2 (-2 k_x^2 \mc{C}^2+\Omega^2)}{\Omega[(k_x^2+K_y^2) \mc{C}^2 -\Omega^2]} 
& 0 
& 0 
& \frac{K_y \mc{C}^2 (K_y^2 (\mc{U}^2-\mc{C}^2)-k_x^2 (\mc{U}^2+\mc{C}^2)+\Omega^2)}{\gamma [(k_x^2+K_y^2) \mc{C}^2-\Omega^2]} 
\medskip\\
0 
& 
0 
& -\frac{K_y^2 \mc{U}^2}{\Omega}+\Omega  
& 0 
& 0 
\medskip\\
\frac{2 k_x K_y^2 \mc{U}^2}{(k_x^2+K_y^2) \mc{C}^2-\Omega^2} 
& \frac{K_y^3 \mc{U}^2}{(k_x^2+K_y^2) \mc{C}^2-\Omega^2} - K_y
& 0 
& -\frac{K_y^2 \mc{U}^2}{\Omega}+\Omega  
& \frac{K_y^2 (k_x^2+K_y^2) \mc{C}^2 \mc{U}^2}{\gamma \Omega [(k_x^2+K_y^2) \mc{C}^2 -\Omega^2]} 
\medskip\\
\frac{2 k_x K_y^2 \mc{U}^2 \gamma}{(k_x^2+K_y^2) \mc{C}^2-\Omega^2} 
& \gamma K_y \left[\frac{K_y^2 \mc{U}^2}{(k_x^2+K_y^2) \mc{C}^2-\Omega^2}-1\right] 
& 0 
& 0 
& \Omega \left[1+\frac{K_y^2 \mc{U}^2}{(k_x^2+K_y^2) \mc{C}^2-\Omega^2}\right] \\
\end{array}
\right),
\notag
\end{gather}
\end{widetext}
where $\mc{C} \doteq c\sqrt{\gamma} = \sqrt{\gamma\bar{P}/\bar{\varrho}}$ is the sound speed. Then, from \Eq{eq:Qdet0}, one can readily see that there is a modulational mode which satisfies
\begin{gather}\label{eq:ts}
 \Omega^2 = K_y^2 \mc{U}^2.
\end{gather}
This is a stable mode, and it is understood as transverse sound wave corresponding to oscillations of $\msf{v}_z$. The remaining modulational modes are studied below.

\subsection{Modulational instabilities}

Assuming the notation
\begin{gather}
w \doteq \Omega/(k_x \mc{U}),
\quad
\kappa \doteq K_y/k_x,
\quad
\vartheta \doteq \mc{U}/\mc{C},
\end{gather}
one can represent \Eq{eq:Qdet0} [with the root \eq{eq:ts} excluded] as follows:
\begin{gather}\label{eq:det}
\sum_{m=0}^4 \alpha_m w^{2m} = 0,
\end{gather}
where the coefficients $\alpha_m$ are given by
\begin{align}
\alpha_0 &= \kappa^4 [-1-3 \vartheta^2+\vartheta^2 (3+\vartheta^2) \kappa^2+(-1+\vartheta^2)^2 \kappa^4], 
\notag\\
\alpha_1 &= -\kappa^2 [1+2 \vartheta^2+(2+\vartheta^2+8 \vartheta^4) \kappa^2\notag\\
& \mbox{}\hspace{3cm}+(1+\vartheta^2) (1+\vartheta^4) \kappa^4], \notag\\
\alpha_2 & = \vartheta^2 [1+(4+5 \vartheta^2) \kappa^2+3 (1+\vartheta^2+\vartheta^4) \kappa^4], \notag\\
\alpha_3 & = -\vartheta^4 [2+3 (1+\vartheta^2) \kappa^2], \notag\\
\alpha_4 & = \vartheta^6.\notag
\end{align}

\subsubsection{Kelvin---Helmholtz instability}
\label{sec:KH}

Let us first consider \Eq{eq:det} at $\vartheta \to 0$, which corresponds to the incompressible-fluid limit ($\gamma \to \infty$). Then, \Eq{eq:det} becomes
\begin{gather}
-w^2 \kappa^2(1 + \kappa^2)^2 - \kappa^4 (1 - \kappa^4) + \mc{O}(\vartheta^2) = 0.
\end{gather}
This leads to
\begin{gather}
 w^2 = - \kappa^2\, \dfrac{1-\kappa^2}{1+\kappa^2} + \mc{O}(\vartheta^2),
 \label{eq:dispersion_relation_c_infinity}
\end{gather}
which implies that there is a MI at ${\kappa^2 < 1}$. The growth rate $\Gamma \doteq \Im\Omega$ of this MI is given~by
\begin{gather}
\Gamma \approx |K_y|\mc{U}\sqrt{\dfrac{k_x^2-K_y^2}{k_x^2+K_y^2}},
\end{gather}
and the maximum rate, $\Gamma_{\rm max} \approx k_x\mc{U}(3-2\sqrt{2})^{1/2}$, corresponds to $\smash{|K_y| = k_x(\sqrt{2}-1)^{1/2}}$. Notably, a similar calculation based on the WKE \eq{eq:wke} [as opposed to the complete WME \eq{eq:WM3} used above] leads to $\Gamma \approx |K_y|\mc{U}$ and thus fails to capture the fact that $\Gamma$ is maximized at finite~$|K_y|$.

\begin{figure}
\begin{center}
\includegraphics[width=.45\textwidth]{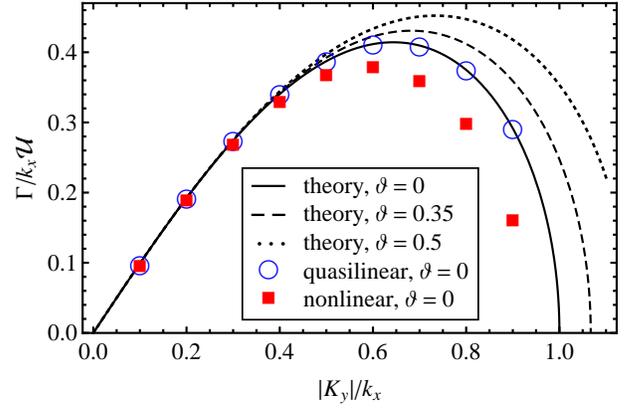}
\caption{The growth rate of the two-dimensional MI of the Kelvin--Helmholtz type (\Sec{sec:KH}) of the equilibrium flow \eq{eq:iniprofile} at small $\vartheta \doteq \mathcal{U}/\mathcal{C}$. The modulation wave vector is given by \Eq{eq:Ky}. The rate $\Gamma$ is measured in units $k_x\mathcal{U}$, the modulation wave number $K_y$ is measured in units~$k_x$. The curves show analytic results [\Eq{eq:lead}] for $\vartheta = 0$ (solid), $\vartheta = 0.35$ (dashed), and $\vartheta = 0.5$ (dotted). The plot markers show the results extrapolated from direct numerical simulations in the incompressible-fluid limit ($\gamma \to \infty$, or $\vartheta \to 0$): quasilinear (blue circles) and nonlinear (red squares).}
\label{fig:gamma1}
\end{center}
\end{figure}

As seen easily, the mode \eq{eq:dispersion_relation_c_infinity} is comprised of $\bar{v}_x$ oscillations, so the MI that we found is in fact a variation of the Kelvin--Helmholtz instability. Note that \Eq{eq:dispersion_relation_c_infinity} coincides, as it should, with the dispersion relation of the so-called zonostrophic instability known from the Hasegawa--Mima model in the appropriate limit \cite{foot:hme}. Including the term $\mc{O}(\vartheta^2)$ perturbatively, we can also extend this dispersion relation to finite sound speed:
\begin{gather}\label{eq:lead}
w^2 = -\kappa^2 \dfrac{1-\kappa^2}{1 + \kappa^2} - \dfrac{4\kappa^6\vartheta^2}{(1+\kappa^2)^3} + \mc{O}(\vartheta^4).
\end{gather}
These results are illustrated in \Fig{fig:gamma1}, which also shows that our theory agrees with direct numerical simulations.

\subsubsection{Modulational instability of supersonic flows}
\label{sec:supersonic}

Although our quasilinear model works best at small $\tilde{v}$ by design, it is also instructive to consider the limit $\vartheta \gg 1$, which corresponds to the initial flow \eq{eq:iniprofile} being hypersonic. In this case, \Eq{eq:det} becomes
\begin{gather}
w^2 \vartheta^2 \left(w^2-\kappa^2\right)^3 + \vartheta^{-2}b^3 + \mc{O}(\vartheta^{-4}) = 0,
\end{gather}
where the coefficient $b = \mc{O}(1)$ is given by
\begin{multline}
b^3 \doteq \kappa^6+\kappa^8-w^2 \kappa^4 (8+\kappa^2)+w^4 \kappa^2 (5+3 \kappa^2)
\\
-w^6 (2+3 \kappa^2).
\end{multline}
This leads to $w = \pm \kappa + e^{2\pi i m/3}|b|\vartheta^{-1/3}$, where $m$ is integer. To the leading order, $b^3 \approx -4\kappa^6$, so the corresponding dispersion relation is
\begin{gather}\label{eq:smallC1}
w \approx \pm \kappa + e^{2\pi i m/3}\left(\frac{|\kappa|}{2\vartheta^2}\right)^{1/3}, \quad m = 0,1,2.
\end{gather}
The amplitudes of the individual components of $\bar{\psi}$ in these oscillations scale as $\bar{v}_x/\bar{v}_y \sim\kappa$, $\bar{n} \sim \bar{\tau}/\gamma \sim \bar{v}_y/\mc{U}$. (In particular, this means that the mode is largely $\bar{v}_y$-polarized at small $\kappa$.) One can also see that one of these modes is unstable and has the following growth rate:
\begin{gather}\label{eq:smallC2}
\Gamma \approx \frac{\sqrt{3}}{2}\,k_x\mathcal{U} \left(\frac{|K_y| \mc{C}^2}{2k_x\mc{U}^2}\right)^{1/3}.
\end{gather}

\begin{figure}[b]
\begin{center}
\includegraphics[width=.48\textwidth]{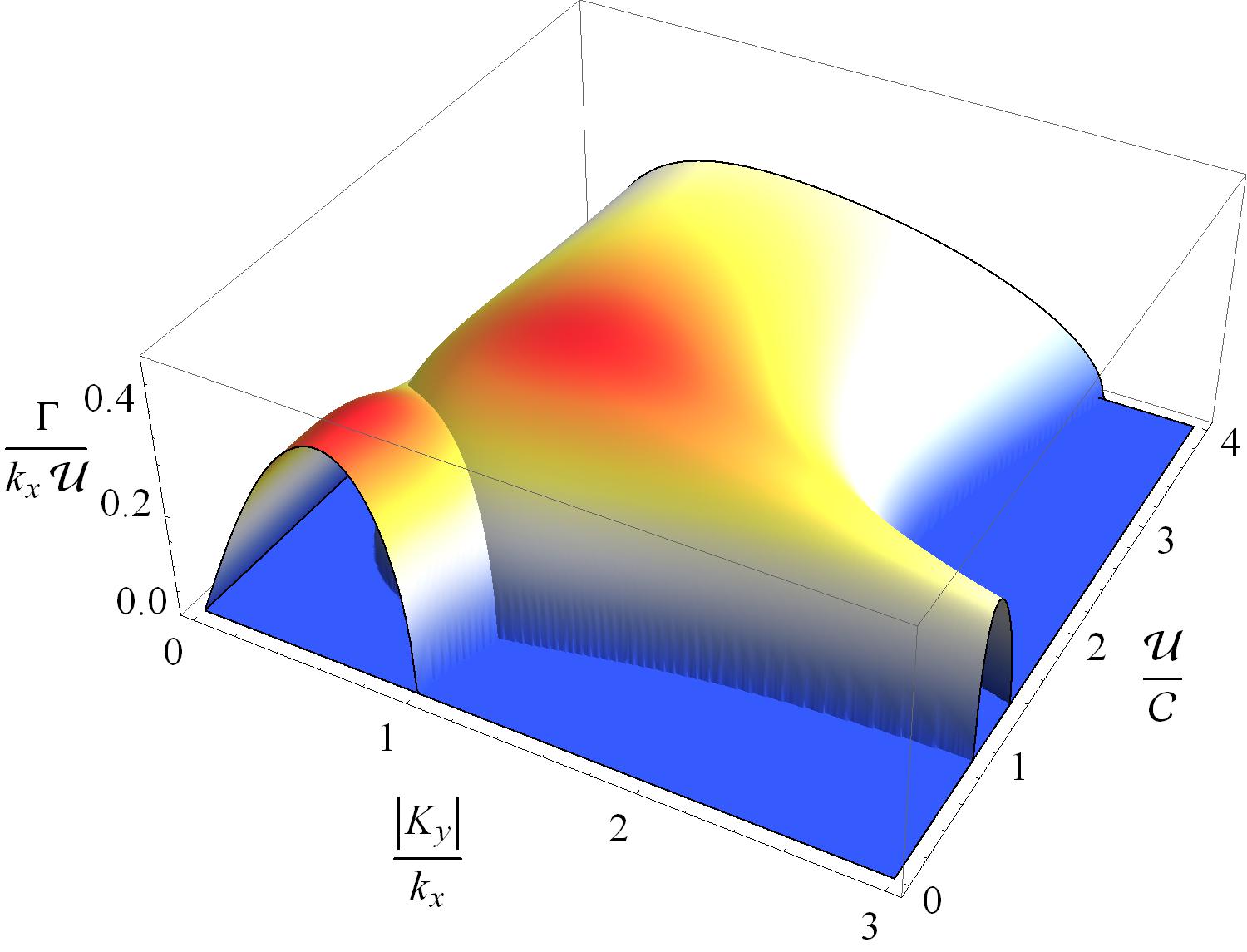}
\caption{The growth rate of a two-dimensional MI of the equilibrium flow \eq{eq:iniprofile} as a function of $\kappa \doteq K_y/k_x$ and $\vartheta \doteq \mc{U}/\mc{C}$. Shown is a numerical solution of \Eq{eq:det}. At $\vartheta \ll 1$, the result is the same as in \Fig{fig:gamma1}. At $\vartheta \gtrsim 1$, an additional branch appears, which corresponds to \Eqs{eq:smallC1} and \eq{eq:smallC2} at $\vartheta \gg 1$.}
\label{fig:gamma2}
\end{center}
\end{figure}

\begin{figure}[t]
\begin{center}
\includegraphics[width=.45\textwidth]{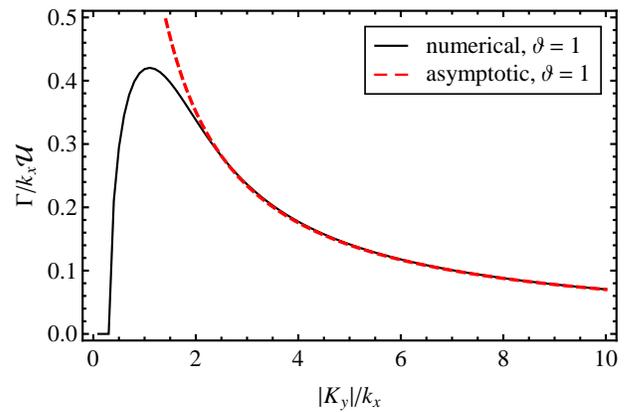}
\caption{Same as in \Fig{fig:gamma2} for $\mathcal{U} = \mathcal{C}$: solid -- numerical solution of \Eq{eq:det}; dashed -- asymptotic formula \eq{eq:asym2}.}
\label{fig:gamma3}
\end{center}
\end{figure}

A numerical solution of the complete dispersion relation \eq{eq:det} for $\Gamma(\kappa, \vartheta)$ is presented in \Fig{fig:gamma2}. It is seen that the second MI survives also for supersonic flows with $\vartheta \sim 1$ [remember that the maximum of $\tilde{v}_x$ is not $\mc{U}$ but $\mc{U}\sqrt{2}$; see \Eq{eq:iniprofile}] but vanishes at smaller~$\vartheta$. At large $\kappa$, the unstable region is localized near $\mc{U} \approx \mc{C}$. A tedious but straightforward calculation shows that in this case,
\begin{gather}\label{eq:asym2}
\Gamma \approx \frac{\sqrt{71}k_x^2\mc{C}}{12K_y},
\end{gather}
where $\vartheta = 1$ and $\kappa \gg 1$ is assumed. [As seen in \Fig{fig:gamma3}, \Eq{eq:asym2} indeed agrees with the full \Eq{eq:det} in the corresponding limit.] Like in \Sec{sec:KH}, this result indicates that $\Gamma$ is maximized at finite~$|K_y|$, which could not have been captured by the GO WKE \eq{eq:wke} but is captured by the `quantumlike' WME \eq{eq:WM3} that we used above. Also note that the absolute maximum of $\Gamma/(k_x \mc{U})$ over all $K_y$ and $\mc{U}$ satisfies
\begin{gather}
\left(\frac{\Gamma}{k_x\mc{U}}\right)_{\rm max} \sim 1.
\end{gather}

\section{Conclusions}
\label{sec:conclusions}

In summary, we study quasilinear modulational dynamics of compressible inviscid Navier--Stokes turbulence using the Wigner--Moyal formulation. This formulation presents the turbulence as effective collisionless quantum plasma where fluctuations serve as particles and coherent flows serve as fields through which these `particles' interact. Unlike in previous applications of the Wigner--Moyal formalism to classical waves, in our case, the fluctuation Hamiltonian is a non-Hermitian five-dimensional matrix operator, so there are multiple modulational modes with nontrivial dispersion. As an example, we derive the dispersion relation of two MIs of compressible shear flows. One of these instabilities is specific to supersonic flows. The other one is a Kelvin--Helmholtz-type instability that is a generalization of the known zonostrophic instability of incompressible fluid. Our work is intended as a stepping stone toward improving the understanding of magnetohydrodynamic turbulence, which can be approached with a similar method.

The work was supported by the U.S. DOE through Contract No. DE-AC02-09CH11466.

\appendix

\section{The Wigner--Weyl transform and some properties of Weyl symbols}
\label{sec:weyl}

Here, we briefly summarize our conventions regarding the Wigner--Weyl transform. (For a more detailed presentation, see, for example \Refs{book:tracy, foot:supp}.) Consider a Hilbert space $\mc{H}^n$ of functions defined on the configurations space $\mathbb{R}^n$. Consider any given operator $\oper{A}$ on $\mc{H}^n$. The Wigner--Weyl transform defines a mapping, or projection, of $\oper{A}$ on the $2n$-dimensional `phase space' $(\mathbf{x}, \mathbf{k})$. Specifically, this projection, also called the Weyl image or the phase-space representation of $\oper{A}$, is given by
\begin{gather}
A\left(\mathbf{x}, \mathbf{k}\right) \doteq \int \mathrm{d}^{n} s\ e^{-i\mathbf{k}\cdot \mathbf{s}} \left\langle \mathbf{x}+\mathbf{s}/2\right\vert \hat{A}\left\vert \mathbf{x}-\mathbf{s}/2\right\rangle.
\label{eq:AppendixA_Weyl_symbol_of_operator}
\end{gather}
It can be shown that
\begin{equation*}
\hat{A}\! =\! \dfrac{1}{(2\pi)^n}\!\int\mathrm{d}^{n}\!x\ \mathrm{d}^{n}\!k\ \mathrm{d}^{n}\!s\  e^{-i\mathbf{k}\cdot\mathbf{s}}A(\mathbf{x, \mathbf{k}})\vert\mathbf{x}-\mathbf{s}/2\rangle\langle\mathbf{x}+\mathbf{s}/2\vert,
\notag
\end{equation*}
which defines the inverse Wigner--Weyl transform. It can also be shown that for any operator $\hat{A}$, its matrix elements in the coordinate representation, $\mc{A}(\mathbf{x}, \mathbf{x}')\doteq\langle\mathbf{x}\vert\hat{A}\vert\mathbf{x}'\rangle$, can be expressed as
\begin{gather}
\mc{A}(\mathbf{x}, \mathbf{x}') = \dfrac{1}{(2\pi)^n}\int\mathrm{d}^{n}\!k\ e^{-i\mathbf{k}\cdot(\mathbf{x}-\mathbf{x}')}A\left( \dfrac{\mathbf{x}+\mathbf{x}'}{2}, \mathbf{k}\right).
\notag
\end{gather}
Thus, $A(\mathbf{x}, \mathbf{k})$ can be interpreted as the spectrum of $\mc{A}(\mathbf{x}, \mathbf{x}')$, and in particular,
\begin{gather}
\mc{A}(\mathbf{x}, \mathbf{x}) = \int \dfrac{\mathrm{d}^{n}\!k}{(2\pi)^n}\,A(\mathbf{x},\mathbf{k}).
\label{eq:AppendixA_Spectrum}
\end{gather}

For any $\hat{C} = \hat{A}\hat{B}$, the corresponding Weyl symbols satisfy
\begin{gather}
C(\mathbf{x},\mathbf{k}) = A(\mathbf{x},\mathbf{k})\star B(\mathbf{x}, \mathbf{k}).
\label{eq:AppendixA_Weyl_symbol_of_C}
\end{gather}
Here, $\star$ is the \textit{Moyal product}, which is given by
\begin{gather}
A(\mathbf{x}, \mathbf{k})\star B(\mathbf{x}, \mathbf{k}) \doteq A(\mathbf{x}, \mathbf{k})e^{i\hat{\mc{L}}/2}B(\mathbf{x}, \mathbf{k}),
\label{eq:AppendixA_Moyal_Star_Product}
\end{gather}
and $\hat{\mc{L}}$ is the \textit{Janus operator}, which is defined as follows:
\begin{gather}
\hat{\mc{L}} \doteq \cev{\partial_{\mathbf{x}}}\cdot\vec{\partial_{\mathbf{k}}} - \cev{\partial_{\mathbf{k}}}\cdot\vec{\partial_{\mathbf{x}}}.
\label{eq:AppendixA_Janus_operator}
\end{gather}
The arrows indicate the directions in which the derivatives act, so for example, $A\hat{\mc{L}}B = \lbrace A, B\rbrace$ is the canonical Poisson bracket,
\begin{gather}
\lbrace A, B\rbrace \doteq \left(\partial_{\mathbf{x}}A\right)\cdot \left(\partial_{\mathbf{k}}B\right) - \left(\partial_{\mathbf{k}}A\right)\cdot\left(\partial_{\mathbf{x}}B\right).
\label{eq:AppendixA_Poisson_bracket}
\end{gather}
In particular, for any constant $\mathbf{K}$, one has
\begin{gather}
    A(\mathbf{k})\star e^{i\mathbf{K}\cdot\mathbf{x}}  = A(\mathbf{k} + \mathbf{K}/2) e^{i\mathbf{K}\cdot\mathbf{x}},
\end{gather}
as seen from the formal Taylor expansion of $A(\mathbf{k} + \mathbf{K}/2)$ in $\mathbf{K}$. Also, for any $A(\mathbf{x}, \mathbf{k})$, one has
\begin{gather}
    k_{a}\star A\left(\mathbf{x},\mathbf{k}\right) = k_a A\left(\mathbf{x},\mathbf{k}\right) 
    - (i/2) \partial_a A\left(\mathbf{x},\mathbf{k}\right).
    \label{eq:AppA_k_star_A}
\end{gather}

Let us also consider several special cases of particular interest. The Weyl symbols of the identity, position, and momentum operators are given by
    \begin{gather}
        \hat{1} \Leftrightarrow 1, 
        \quad 
        \hat{x}_a \Leftrightarrow x_a, 
        \quad 
        \hat{k}_a \Leftrightarrow k_a.
    \end{gather}
(Here $\Leftrightarrow$ denotes the correspondence between operators and their Weyl symbols.) Also, for any given functions $f$ and $g$, one has
\begin{gather}
f(\hat{\mathbf{x}}) \Leftrightarrow f(\mathbf{x}),
\quad  
g(\hat{\mathbf{k}}) \Leftrightarrow g(\mathbf{k}),
\label{eq:AppendixA_Weyl_transform_of_f(x)k}
\end{gather}
and similarly, using \Eq{eq:AppendixA_Moyal_Star_Product}, one obtains
\begin{subequations}
\begin{align}
f(\hat{\mathbf{x}})\hat{k}_a & \Leftrightarrow k_a f(\mathbf{x}) + (i/2)\,\partial_a f(\mathbf{x}), \\
\hat{k}_a f(\hat{\mathbf{x}}) & \Leftrightarrow k_a f(\mathbf{x}) - (i/2)\,\partial_a f(\mathbf{x}),
\label{eq:AppendixA_Weyl_transform_of_kf(x)}
\end{align}
\end{subequations}
where the latter also flows from \Eq{eq:AppA_k_star_A}.

\mbox{}

\section{Explicit formulas for the Hamiltonians and their symbols}
\label{app:matrices}

Here, we present explicit formulas for the Hamiltonians used in the main text and also for their symbols. In particular, the matrix function $H_0$ [\Eq{eq:H0}] is given by
\begin{gather}
    H_0(\mathbf{k}) \!=\! \begin{pmatrix}
    0  &  0  &  0  &  0  &  c^2 k_x  \\
    0  &  0  &  0  &  0  &  c^2 k_y  \\
    0  &  0  &  0  &  0  &  c^2 k_z  \\
    k_x  &  k_y  &  k_z  &  0  &  0  \\
    \gamma k_x  &  \gamma k_y  &  \gamma k_z  &  0  &  0 
    \end{pmatrix}.
\label{eq:AppB_H0}
\end{gather}
The operator $\hat{H}_{\rm int}$ is given by
\begin{widetext}
\begin{gather}
    \hat{H}_{\rm int} \!=\! \begin{pmatrix}
    \bar{\mathbf{v}}\!\cdot\!\hat{\mathbf{k}} - i\partial_x \bar{v}_x  &  -i\partial_y \bar{v}_x  &  -i\partial_z \bar{v}_x  &  ic^2(1+\bar{n})^{-2} (\partial_x \bar{\tau})  &  -c^2 \bar{n}(1+\bar{n})^{-1}\hat{k}_x \\
    -i\partial_x \bar{v}_y  &  \bar{\mathbf{v}}\!\cdot\!\hat{\mathbf{k}} -i\partial_y \bar{v}_y  &  -i\partial_z \bar{v}_y  &  ic^2(1+\bar{n})^{-2}(\partial_y \bar{\tau})  &  -c^2 \bar{n}(1+\bar{n})^{-1}\hat{k}_y \\
    -i\partial_x \bar{v}_z  &  -i\partial_y \bar{v}_z  &  \bar{\mathbf{v}}\!\cdot\!\hat{\mathbf{k}} -i\partial_z \bar{v}_z  &  ic^2(1+\bar{n})^{-2}(\partial_z \bar{\tau})  &  -c^2\bar{n}(1+\bar{n})^{-1}\hat{k}_z \\
    \bar{n}\hat{k}_x - i\partial_x \bar{n}  &  \bar{n}\hat{k}_y - i\partial_y \bar{n}  &  \bar{n}\hat{k}_z - i\partial_z \bar{n}  &  \bar{\mathbf{v}}\!\cdot\!\hat{\mathbf{k}} - i\nabla\!\cdot\!\bar{\mathbf{v}}  &  0  \\
    \gamma\bar{\tau}\hat{k}_x - i\partial_x \bar{\tau}  &  \gamma\bar{\tau}\hat{k}_y -i\partial_y \bar{\tau}  & \gamma\bar{\tau}\hat{k}_z - i\partial_z \bar{\tau}  &  0  &  \bar{\mathbf{v}}\!\cdot\!\hat{\mathbf{k}} - i\gamma\nabla\!\cdot\!\bar{\mathbf{v}}
    \end{pmatrix}.
    \label{eq:AppB_Hint}
\end{gather}
The corresponding Weyl symbol is as follows:
\begin{subequations}\label{eq:Hintsymbol}
\begin{gather}
    H_{\rm int}(t, \mathbf{x}, \mathbf{k}) \!= \begin{pmatrix}
    \bar{\mathbf{v}}\!\cdot\!\mathbf{k} - i\partial_x \bar{v}_x  &  -i\partial_y \bar{v}_x  &  -i\partial_z \bar{v}_x  &  ic^2(1+\bar{n})^{-2} (\partial_x \bar{\tau})  &  - c^2 \bar{n}(1+\bar{n})^{-1} k_x \\
    -i\partial_x \bar{v}_y  &  \bar{\mathbf{v}}\!\cdot\!\mathbf{k} -i\partial_y \bar{v}_y  &  -i\partial_z \bar{v}_y  &  ic^2(1+\bar{n})^{-2}(\partial_y \bar{\tau})  &  -c^2 \bar{n}(1+\bar{n})^{-1} k_y \\
    -i\partial_x \bar{v}_z  &  -i\partial_y \bar{v}_z  &  \bar{\mathbf{v}}\!\cdot\!\mathbf{k} -i\partial_z \bar{v}_z  &  ic^2(1+\bar{n})^{-2}(\partial_z \bar{\tau})  &  -c^2\bar{n}(1+\bar{n})^{-1} k_z \\
    \bar{n} k_x - i\partial_x \bar{n}  &  \bar{n} k_y - i\partial_y \bar{n}  &  \bar{n} k_z - i\partial_z \bar{n}  &  \bar{\mathbf{v}}\!\cdot\!\mathbf{k} - i\nabla\!\cdot\!\bar{\mathbf{v}}  &  0  \\
    \gamma\bar{\tau} k_x - i\partial_x \bar{\tau}  &  \gamma\bar{\tau} k_y -i\partial_y \bar{\tau}  & \gamma\bar{\tau} k_z - i\partial_z \bar{\tau}  &  0  & \bar{\mathbf{v}}\!\cdot\!\mathbf{k} - i\gamma\nabla\!\cdot\!\bar{\mathbf{v}}
    \end{pmatrix} +\dfrac{i}{2}\,\Delta H_{\rm int},
    \\
    \Delta H_{\rm int}(t, \mathbf{x}, \mathbf{k}) = 
    \begin{pmatrix}
    \nabla\!\cdot\!\bar{\mathbf{v}} &  0  &  0 &  0  & -c^2(1+\bar{n})^{-2}(\partial_x \bar{n}) \\
    0  &  \nabla\!\cdot\!\bar{\mathbf{v}}  &  0  &  0  &  -c^2(1+\bar{n})^{-2}(\partial_y \bar{n}) \\
    0  &  0  &  \nabla\!\cdot\!\bar{\mathbf{v}}  &  0  &  -c^2(1+\bar{n})^{-2}(\partial_z \bar{n})  \\
    \partial_x \bar{n}  &  \partial_y \bar{n}  &  \partial_z \bar{n}  &  \nabla\!\cdot\!\bar{\mathbf{v}}  &  0  \\
    \gamma\partial_x \bar{\tau}  &  \gamma \partial_y \bar{\tau}  &  \gamma \partial_z \bar{\tau}  &  0  &  \nabla\!\cdot\!\bar{\mathbf{v}}
    \end{pmatrix}.
\end{gather}
\end{subequations}
The linearized symbol $h$ is obtained from this by neglecting $\bar{n}$ in $1 + \bar{n}$. Hence, $\msf{h}^{(+)}$ [\Eq{eq:H_h1_with_F_and_G}] is given by
\begin{subequations}\label{eq:AppB_h+}
\begin{gather}
    \msf{h}^{(+)}(\mathbf{K}, \mathbf{k}, \bar{\psi}) = \begin{pmatrix}
    \boldsymbol{\msf{v}}\!\cdot\!\mathbf{k} + \msf{v}_xK_x  &  \msf{v}_xK_y  &  \msf{v}_x K_z  &  -c^2 \uptau K_x  &  -c^2\msf{n}k_x \\
    \msf{v}_y K_x  &  \boldsymbol{\msf{v}}\!\cdot\!\mathbf{k} + \msf{v}_y K_y  &  \msf{v}_yK_z  &  -c^2\uptau K_y  &  -c^2\msf{n}k_y  \\
    \msf{v}_z K_x  &  \msf{v}_z K_y  &  \boldsymbol{\msf{v}}\!\cdot\!\mathbf{k}+\msf{v}_z K_z  &  -c^2\uptau K_z  &  -c^2\msf{n}k_z \\
    \msf{n}(k_x + K_x)  &  \msf{n}(k_y + K_y)  &  \msf{n}(k_z + K_z)  &  \boldsymbol{\msf{v}}\!\cdot\!(\mathbf{k}+\mathbf{K})  &  0  \\
    \uptau (\gamma k_x + K_x)  &  \uptau (\gamma k_y + K_y)  &  \uptau (\gamma k_z + K_z)  &  0  &  \boldsymbol{\msf{v}}\!\cdot\!(\mathbf{k} + \gamma \mathbf{K})
    \end{pmatrix}-\dfrac{1}{2}\,\Delta\msf{h}^{(+)}(\mathbf{K}, \mathbf{k}, \bar{\psi}),
    \\
    \Delta\msf{h}^{(+)}(\mathbf{K}, \mathbf{k}, \bar{\psi}) =
    \begin{pmatrix}
    \boldsymbol{\msf{v}}\!\cdot\!\mathbf{K}  &  0  &  0  &  0  &  -c^2\msf{n}K_x \\
    0  &  \boldsymbol{\msf{v}}\!\cdot\!\mathbf{K}  &  0  &  0  &  -c^2\msf{n}K_y \\
    0  &  0  &  \boldsymbol{\msf{v}}\!\cdot\!\mathbf{K}  &  0  &  -c^2\msf{n}K_z \\
    \msf{n}K_x  &  \msf{n}K_y  &  \msf{n}K_z  & \boldsymbol{\msf{v}}\!\cdot\!\mathbf{K}  &  0  \\
    \gamma \uptau K_x  &  \gamma \uptau K_y  &  \gamma \uptau K_z  &  0  &  \boldsymbol{\msf{v}}\!\cdot\!\mathbf{K}
    \end{pmatrix}.
\end{gather}
\end{subequations}
Clearly, $\msf{h}^{(-)\dag}(\mathbf{K}, \mathbf{k}, \bar{\psi}) = \msf{h}^{(+)}(-\mathbf{K}, \mathbf{k}, \bar{\psi}^*)$. Assuming that $\mathbf{K}$ is real, this leads to $\msf{h}^{(-)}(\mathbf{K}, \mathbf{k}, \bar{\psi}) = [\msf{h}^{(+)}(-\mathbf{K}, \mathbf{k}, \bar{\psi})]^\intercal$ (here $^\intercal$ denotes transposition), or explicitly,
\begin{subequations}\label{eq:AppB_h-}
\begin{gather}
        \msf{h}^{(-)}(\mathbf{K}, \mathbf{k}, \bar{\psi}) = \begin{pmatrix}
        \boldsymbol{\msf{v}}\!\cdot\!\mathbf{k} - \msf{v}_x K_x  &  -\msf{v}_y K_x  &  -\msf{v}_z K_x  &  \msf{n}(k_x-K_x)  &  \gamma\uptau(k_x - K_x) \\
        -\msf{v}_x K_y  &  \boldsymbol{\msf{v}}\!\cdot\!\mathbf{k} - \msf{v}_y K_y  &  -\msf{v}_z K_y  &  \msf{n}(k_y-K_y)  &  \gamma \uptau (k_y-K_y) \\
        -\msf{v}_x K_z  &  - \msf{v}_y K_z  &  \boldsymbol{\msf{v}}\!\cdot\!\mathbf{k}-\msf{v}_z K_z  &  \msf{n}(k_z - K_z)  &  \gamma \uptau (k_z - K_z)  \\
        c^2\uptau K_x  &  c^2\uptau K_y  &  c^2 \uptau K_z  &  \boldsymbol{\msf{v}}\!\cdot\!(\mathbf{k}-\mathbf{K})  &  0  \\
        -c^2 \msf{n}k_x  &  -c^2\msf{n}k_y  &  -c^2\msf{n}k_z  &  0  &  \boldsymbol{\msf{v}}\!\cdot\!(\mathbf{k} - \gamma \mathbf{K})
        \end{pmatrix} +\dfrac{1}{2}\,\Delta \msf{h}^{(-)}(\mathbf{K}, \mathbf{k}, \bar{\psi}),
            \\
\Delta \msf{h}^{(-)}(\mathbf{K}, \mathbf{k}, \bar{\psi})=    
        \begin{pmatrix}
        \boldsymbol{\msf{v}}\!\cdot\!\mathbf{K}  &  0  &  0  &  \msf{n}K_x  &  \gamma \uptau K_x  \\
        0  &  \boldsymbol{\msf{v}}\!\cdot\!\mathbf{K}  &  0  &  \msf{n}K_y  &  \gamma \uptau K_y  \\
        0  &  0  &  \boldsymbol{\msf{v}}\!\cdot\!\mathbf{K}  &  \msf{n}K_z  &  \gamma \uptau K_z  \\
        0  &  0  &  0  &  \boldsymbol{\msf{v}}\!\cdot\!\mathbf{K}  &  0  \\
        -c^2 \msf{n}K_x  &  -c^2 \msf{n}K_y  &  -c^2\msf{n}K_z  &  0  &  \boldsymbol{\msf{v}}\!\cdot\!\mathbf{K}
        \end{pmatrix}.
\end{gather}
\end{subequations}
\end{widetext}

\bibliography{my,main,turbulence,foot}

\begin{thebibliography}{10}
\bibitem{foot:soliton}
A simple example is the generation of solitary structures via MIs. The soliton
  size is determined by the balance between nonlinearity and diffraction, so
  solitons cannot be described within the GO approximation regardless of how
  large their length is \cite{my:soliton}.
\bibitem{foot:hme}
For example, see Eq.~(2.12) in \Ref{my:ppo}. The appropriate limit corresponds
  to $\beta \to 0$ and $L_D \to \infty$, assuming the notation used in the
  mentioned paper. Also, $2 p \Phi_0$ there coincides with our $\mc{U}\sqrt{2}$
  by definition, $p$ is the same as our $k_x$, and $q$ is the same as our
  $K_y$.
\end{thebibliography}

\end{document}